\documentclass[runningheads]{llncs}
\usepackage[numbers,sort&compress]{natbib}

\usepackage{graphicx}

\newcommand{\searchservice}[0]{{\sc RoutIR}}
\newcommand{\routir}[0]{{\sc RoutIR}}

\usepackage{booktabs}
\usepackage{hyperref}
\usepackage{xcolor}
\usepackage{listings} 
\definecolor{deepblue}{rgb}{0,0,0.5}
\definecolor{deepred}{rgb}{0.6,0,0}
\definecolor{deepgreen}{rgb}{0,0.5,0}
\definecolor{codegray}{rgb}{0.5,0.5,0.5}
\definecolor{codebackground}{rgb}{0.97,0.97,0.97}

\newcommand\pythonstyle{\lstset{
    language=Python,
    basicstyle=\ttfamily\footnotesize,
    morekeywords={self},
    keywordstyle=\bfseries\color{deepblue},
    emphstyle=\bfseries\color{deepgreen},
    stringstyle=\color{deepred},
    commentstyle=\color{codegray}\itshape,
    frame=tb,
    showstringspaces=false,
    numbers=left,
    numberstyle=\tiny\color{codegray},
    numbersep=8pt,
    tabsize=4,
    breaklines=true,
    breakatwhitespace=false,
    backgroundcolor=\color{codebackground}
}}

\lstdefinelanguage{json}{
    basicstyle=\ttfamily\footnotesize,
    morestring=[b]",
    morestring=[d]',
    stringstyle=\color{deepred},
    comment=[l]{//},
    morecomment=[s]{/*}{*/},
    commentstyle=\color{codegray}\itshape,
    literate=
        *{:}{{{\color{deepblue}{:}}}}{1}
        {,}{{{\color{deepblue}{,}}}}{1}
        {\{}{{{\color{deepblue}{\{}}}}{1}
        {\}}{{{\color{deepblue}{\}}}}}{1}
        {[}{{{\color{deepblue}{[}}}}{1}
        {]}{{{\color{deepblue}{]}}}}{1}
        {true}{{{\color{deepgreen}{true}}}}{4}
        {false}{{{\color{deepgreen}{false}}}}{5}
        {null}{{{\color{deepgreen}{null}}}}{4},
}

\newcommand\jsonstyle{\lstset{
    language=json,
    frame=tb,
    showstringspaces=false,
    numbers=left,
    numberstyle=\tiny\color{codegray},
    numbersep=8pt,
    tabsize=2,
    breaklines=true,
    breakatwhitespace=false,
    backgroundcolor=\color{codebackground}
}}

\newcommand\bashstyle{\lstset{
    language=bash,
    basicstyle=\ttfamily\footnotesize,
    keywordstyle=\bfseries\color{deepblue},
    stringstyle=\color{deepred},
    commentstyle=\color{codegray}\itshape,
    emphstyle=\bfseries\color{deepgreen},
    frame=tb,
    showstringspaces=false,
    numbers=left,
    numberstyle=\tiny\color{codegray},
    numbersep=8pt,
    tabsize=4,
    breaklines=true,
    breakatwhitespace=false,
    backgroundcolor=\color{codebackground},
    morekeywords={sudo,apt-get,pip,virtualenv,source},
    alsoletter={-}
}}

\lstnewenvironment{python}[1][]{\pythonstyle\lstset{#1}}{}
\lstnewenvironment{json}[1][]{\jsonstyle\lstset{#1}}{}
\lstnewenvironment{bash}[1][]{\bashstyle\lstset{#1}}{}

\newcommand{\pythoninline}[1]{{\pythonstyle\lstinline!#1!}}
\newcommand{\jsoninline}[1]{{\jsonstyle\lstinline!#1!}}
\newcommand{\bashinline}[1]{{\bashstyle\lstinline!#1!}}

\usepackage{todonotes}

\usepackage{orcidlink}
\newcommand\orcid[1]{\textsuperscript{\orcidlink{#1}}}

\begin{document}
\title{\searchservice{}: Fast Serving of Retrieval Pipelines for Retrieval-Augmented Generation}
\titlerunning{\searchservice{}: Fast Serving of Retrieval Pipelines for RAG}

\author{Eugene Yang\inst{1}\orcid{0000-0002-0051-1535} \and 
Andrew Yates\inst{1}\orcid{0000-0002-5970-880X} \and 
Dawn Lawrie\inst{1}\orcid{0000-0001-7347-7086} \and \\
James Mayfield\inst{1}\orcid{0000-0003-3866-3013} \and
Trevor Adriaanse\inst{2}\orcid{0009-0004-6430-2320}
}

\authorrunning{E. Yang et al.}
\institute{Human Language Technology Center of Excellence, Johns Hopkins University, Baltimore, MD 21211, USA \\
\email{\{eugene.yang,andrew.yates,lawrie,mayfield\}@jhu.edu}
\and 
Johns Hopkins University, Baltimore, MD 21211, USA \\
\email{tadriaa1@jhu.edu}
}

\maketitle              %
\begin{abstract}
Retrieval models are key components of Retrieval-Augmented Generation (RAG) systems, which generate search queries, process the documents returned, and generate a response.
RAG systems are often dynamic and may involve multiple rounds of retrieval.
While many state-of-the-art retrieval methods are available through academic IR platforms, these platforms are typically designed for the Cranfield paradigm in which all queries are known up front and can be batch processed offline.
This simplification accelerates research
but leaves state-of-the-art retrieval models unable to support downstream applications that require online services, such as arbitrary dynamic RAG pipelines that involve looping, feedback, or even self-organizing agents. 
In this work, we introduce \searchservice{},
a Python package that provides a simple and efficient HTTP API that wraps arbitrary retrieval methods,
including first stage retrieval, reranking, query expansion, and result fusion.
By providing a minimal JSON configuration file specifying the retrieval models to serve, \routir{} can be used to construct and query retrieval pipelines on-the-fly using any permutation of available models (e.g., fusing the results of several first-stage retrieval methods followed by reranking).
The API automatically performs asynchronous query batching and caches results by default.
While many state-of-the-art retrieval methods are already supported by the package,
\searchservice{} is also easily expandable by implementing the \textit{Engine} abstract class. 
The package is open-sourced and publicly available on GitHub: \url{http://github.com/hltcoe/routir}. 

\end{abstract}

\section{Introduction}

\begin{figure}[t]
    \centering
    \includegraphics[width=\linewidth]{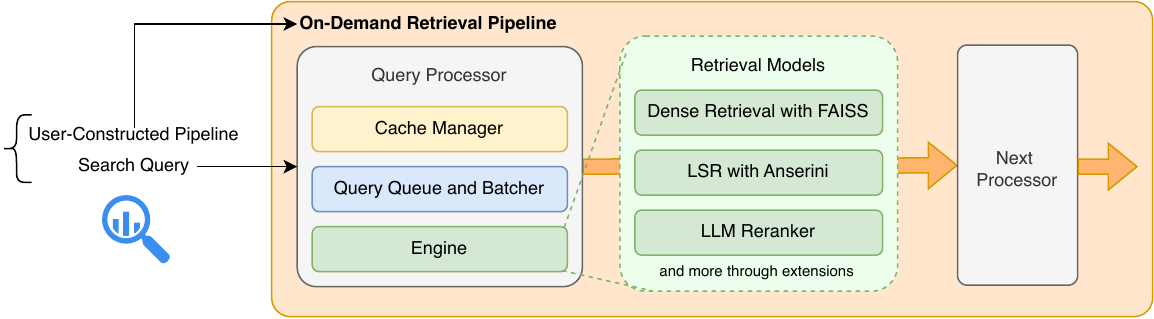}
    \caption{Service architecture of \routir.
    The user's HTTP request specifies a retrieval pipeline and search query, illustrated on the left.
    \routir{} orchestrates the retrieval pipeline on the fly and processes the query with each query processor.
    Each processor manages queuing, caching, and query processing.
    The ``Next Processor'' enables, for example, reranking or result fusion.
    See Figure~\ref{fig:pipeline-request} for a full JSON example of a request.}
    \label{fig:architecture}
\end{figure}

Information retrieval research typically requires system comparison
using a fixed set of queries and documents, following the Cranfield paradigm~\cite{cranfield}. 
Such experiments can be conducted through either batched or sequential query processing,
depending on whether measuring query latency is critical to the study. 
Regardless, the queries are predefined,
so the experimental environment can be static and read directly from a file.
However, embedding retrieval models in a larger dynamic system pipeline creates challenges and overhead when conducting experiments. 

Particularly in Retrieval-Augmented Generation (RAG),
the primary pipeline involves one or more large language models (LLMs)~\cite{xu2025collabrag, dong2025rag}
that generate search queries, digest retrieved documents, and draft the response. 
Instead of a linear process such as Fusion-in-Decoder~\cite{izacard2021leveraging}
or GINGER~\cite{lajewska2025ginger},
RAG systems are increasingly complex and dynamic,
with multiple rounds of retrieval~\cite{dong2025rag, fan2025kirag, shao2024assisting, asai2024self, yan2024corrective}. 
A static experimental environment is not sufficient for these systems,
because the queries are not known up front.
Tools for offline query processing need to be modified to include the generative components
or wrapped to provide an API that can accommodate online querying.
However, wrapping offline retrieval pipelines with reasonable query and resource efficiency is non-trivial, and this task is orthogonal to implementing a retrieval model.
We introduce an open-source package, \routir{},
that allows users to provide pipelines composed of the latest retrieval models
as an HTTP API for use with downstream applications like dynamic RAG systems.

\routir{} supports (1) simple service configuration for common model architectures without additional coding;
(2) minimal wrappers for incorporating new models;
(3) dynamic batching and queuing for concurrent and asynchronous requests;
(4) fast and robust caching in memory and Redis;\footnote{\url{https://redis.io/}}
and (5) reliable and easy-to-use HTTP API endpoints that do not require additional client-side packages. 
Such capabilities provide fast and reproducible experimental environments for RAG research,
while enabling new retrieval models to be incorporated easily with minimal engineering overhead. 
While \routir{} is designed primarily for academic research, which does not implement various security measures for an API service, it can be used for internal prototyping in industry to spin up a proof-of-concept application. 

\routir{} has been deployed in various research settings where it has been demonstrated to be robust and reliable. 
During JHU SCALE 2025,\footnote{\url{https://hltcoe.jhu.edu/research/scale/scale-2025/}}
a ten-week research workshop at Johns Hopkins University attended by more than 50 researchers working on long-form RAG,
\routir{} was able to provide the retrieval API for PLAID-X~\cite{yang2024translate},
SPLADE-v3~\cite{lassance2024spladev3newbaselinessplade},
and Qwen3 Embedding~\cite{zhang2025qwen3} on the TREC NeuCLIR~\cite{lawrie2024overview},
TREC RAGTIME~\cite{lawrie2025overview},
TREC BioGen~\cite{biogen2025overview},
and TREC RAG (i.e., MS-MARCO v2 Passage)~\cite{rag2025overview} document collections
using only three NVIDIA 24GB TITAN RTXs. 
In this use case, all three retrieval systems were able to provide search results
with a reasonable latency without caching,
and to provide results nearly instantaneously when results were cached.
With asynchronous HTTP requests, which is the typical use case in RAG research,
since multiple queries are usually generated and searched at the same time,
\routir{} provides a throughput of 3 to 10 queries per second depending on the underlying model. 
\routir{} also powered the search service for the TREC RAGTIME track\footnote{\url{https://trec-ragtime.github.io/search\_api.html}}
to provide the same PLAID-X model for all track participants.
With only CPU resources, the endpoint provides a latency of around 600 ms on an AWS virtual machine
when serving both TREC NeuCLIR and RAGTIME collections. 

In this paper, we document the design decisions, architecture, and several use cases for \routir. 
The package is publicly available on PyPI with implementation available on GitHub.

\section{Related Work}

\paragraph{Academic IR platforms} have a long history going back to at least the mid-1980s~\cite{buckley1985implementation}.
Platforms like
Galago~\cite{cartright2012galago},
Indri~\cite{strohman2005indri},
Patapsco~\cite{patapsco}, and
Terrier~\cite{OunisTerrierA} support offline experiments with traditional statistical methods,
while platforms like
Anserini~\cite{anserini,pyserini},
Capreolus~\cite{yates2020capreolus,yates2020flexible},
Experimaestro~\cite{piwowarski2020experimaestro},
FlexNeuART~\cite{boytsov2020flexible},
LLM4Ranking~\cite{liu2025llm4rankingeasytouseframeworkutilizing},
LLM-Rankers~\cite{zhuang2024setwise},
OpenNIR~\cite{macavaney2020opennir},
PyTerrier~\cite{PyTerrier},
RankLLM~\cite{zhuang2024setwise}, and
Tevatron~\cite{Gao2022TevatronAE,ma2025tevatron}
provide support for modern neural first-stage retrieval and reranking methods~\cite{lin2022pretrained}.
These platforms are typically designed for offline IR experiments following the Cranfield paradigm.
They provide reproducible indexing and searching capabilities to support experimentation
where test queries are predefined and can be processed sequentially or in batch. 
Rather than providing general-purpose toolkits,
there has been a recent trend towards streamlining experimentation
using tools designed to run specific benchmarks with a predefined set of collections,
such as MTEB~\cite{muennighoff-etal-2023-mteb}
and BEIR~\cite{thakur2021beirheterogenousbenchmarkzeroshot}.
While these academic tools are useful for offline retrieval evaluation and have greatly advanced the field,
their focus on offline usage limits their ability to be embedded in larger systems.
They typically do not provide Web APIs to interface with downstream applications.

This limitation particularly affects RAG,
where retrieval is usually an upstream or interleaving component
providing retrieved information to generative models~\cite{xu2025collabrag, wang2025retrieval, dong2025rag, fan2025kirag, shao2024assisting, asai2024self, yan2024corrective}.
This is a different setting from offline experimentation where the queries are fixed.
RAG pipelines like Open Deep Research\footnote{\url{https://github.com/langchain-ai/open_deep_research}}
can dynamically generate queries to incrementally gather information for generation.
There are two ways to support such iterative RAG pipelines:
(re-)implement the generative component within the IR platform
or provide an API that can be dynamically queried by an external generative library.
PyTerrier-RAG~\cite{macdonald2025constructing} takes the former approach,
whereas \routir{} takes the latter by dynamically accepting and serving queries using state-of-the-art IR methods,
including those implemented by platforms designed for offline use.
Rather than reimplementing the underlying retrieval methods and RAG workflows,
\routir{} provides a streamlined approach to wrap existing methods,
compose them into retrieval pipelines,
and query them online through a HTTP API.
This allows \routir{} to serve the fast-growing RAG community by seamlessly embedding retrieval models into RAG platforms instead of requiring that RAG methods be reimplemented within an academic IR platform.

\paragraph{Production search platforms} like
ElasticSearch\footnote{\url{https://www.elastic.co/elasticsearch}},
OpenSearch\footnote{\url{https://opensearch.org/}}, and
Vespa\footnote{\url{https://vespa.ai/}}
provide HTTP APIs for first-stage retrieval and reranking over collections that have been indexed by those platforms.
These tools are complex, closely integrated with their underlying inverted index and vector database data structures,
and optimized for production use.
These design choices make extending the platforms with new methods non-trivial.
In contrast, \routir{} provides a simple API and flexible method classes that make adding new retrieval models straightforward.

\section{\routir{} Architecture}

\routir{} is a thin, robust wrapper around retrieval models
that provides online service capabilities that are orthogonal to the retrieval models themselves. 
The end user submits an HTTP request to a \routir{} endpoint for each search query,
e.g., \jsoninline{\{"service": "qwen3-neuclir", "query": "where is machu picchu", "limit": 15\}},
and receives the retrieval results of the requested 15 documents in a dictionary of document IDs and scores.
A separate request delivers the document data associated with the document ID.

The core design principle of \routir{} is to be as lightweight as possible
while providing a flexible service layer
to mitigate the overhead of serving state-of-the-art models when they are released. 
In \routir{}, we limit dependencies to only essential packages
and leave model-specific ones, such as
Huggingface Transformers~\cite{wolf2020huggingfacestransformersstateoftheartnatural} and
PyJNIus,\footnote{\url{https://github.com/kivy/pyjnius}}
as extras for users to install when they are needed. 

In this section, we describe the \routir{} service architecture, which is illustrated in Figure~\ref{fig:architecture}. 
\routir{} has three main components: Engines, Processors, and Pipelines. 
\begin{itemize}
    \item \textit{Engines} provide one or more core retrieval capabilities:
    first-stage retrieval using an index, reranking, query rewriting, and result fusion.
    To add new methods to \routir{}, the user writes an Engine subclass that may simply wrap an existing implementation.
    A special \textit{Relay} Engine can be used to access Engines provided by another node running \routir{}.
    \item \textit{Processors} receive search queries as input and perform actions before passing the queries to an Engine.
    By default, they are used to cache results and to batch queries that arrive in quick succession.
    \item \textit{Pipelines} describe how engines are composed to produce a ranking.
    For example, a pipeline could indicate that results from two first-stage retrieval engines
    should be fused and then reranked.
    Pipelines can be composed of available Engines on the fly.
\end{itemize}

\subsection{Retrieval Engines}

Each retrieval model or system should be wrapped as an Engine,
which implements the interface for serving queries.
Each engine can provide one or more of the four core retrieval capabilities:
(a) index searching, (b) query-passage scoring (for reranking), (c) query rewriting (or generation), and (d) result fusion. 

Allowing each engine instance to provide multiple capabilities can minimize the memory footprint when different tasks share the same underlying models. 
For instance, most bi-encoder models can provide both first-stage retrieval that results in a ranked list of documents,
and query-passage scoring (reranking) that enables fast passage selection when 
a RAG pipeline~\cite{guo2025dynamic, cheng2024xrag, jeong2025ecorag} needs to compress the document input
to the most relevant passages to feed to a downstream generation step.
To improve retrieval effectiveness, an Engine can provide a reranker, such as monoT5~\cite{pradeep2021expandomonoduodesignpatterntext}, Qwen3 Reranker~\cite{zhang2025qwen3}, or Rank1~\cite{weller2025rank1testtimecomputereranking}.
Rerankers take a query and a list of strings (e.g. documents) as input and output a ranking over the input list. 
In \routir, modules can be initiated in Python as a standalone model instance
to provide a unified interface;
this is similar to what PyTerrier extensions such as \texttt{pyterrier-colbert} provide. 
However, the primary benefit of this thin wrapper is that it provides
a robust and flexible online search service. 

Queries as inputs to the Engine are batched
(as detailed in Section~\ref{sec:routir:processor})
to provide better service throughput;
this is helpful because most bi-encoders and cross-encoders can score multiple queries and documents
in the same matrix multiplication,
leading to better GPU utilization. 
The common measurement for query-time efficiency has been query latency
measured by sequentially issuing queries to the model
during offline evaluation~\cite{schurman2009performance}.
However, when serving multiple users or queries as a service,
\routir{} optimizes for throughput
(i.e., the number of queries served in a fixed period of time)
since queries can often be served asynchronously. 
Especially for a RAG pipeline that issues multiple queries simultaneously~\cite{duh2025hltcoe-liverag, yang2025hltcoetrec},
all retrieval results must arrive before generation begins;
this suggests the need for high-throughput rather than low-latency
(although the two qualities are usually correlated). 

\subsubsection{Multi-Server Request Routing.}

A special type of Engine is a \textit{Relay} --
an Engine that relays requests to another \routir{} endpoint.
This capability is particularly useful when computing resources are divided among multiple machines
or if some compute nodes are not exposed to a public IP
(a common setup in academic research clusters). 
This is similar to the proxy service in \texttt{LiteLLM}\footnote{\url{https://github.com/BerriAI/litellm}}
that relays LLM requests to compute endpoints without exposing multiple machines to the end users. 
While \routir{} does not offer load-balancing at the request level (which may be included in future versions), it offers triage at the model level to direct requests to models with different resource requirements to different machines. 
\routir{} also supports importing services from a list of endpoints
to simplify configuration
(more on this in Section~\ref{sec:usage:config}). 
This feature provides the backbone for collectively serving multiple retrieval models with one endpoint
in a distributed computing environment,
which is crucial to facilitate complex retrieval pipelines
(more on this later in this section). 

\subsection{Query Processor}\label{sec:routir:processor}

Each Engine is further wrapped in a Processor class,
which handles caching and queuing of input search queries. 
When the processor receives a query,
it is added to the service queue for batching. 
The queue dispatches a batch of queries to the engine
whenever the batch is full (size configurable)
or the maximum wait time is reached (typically 50 to 100 ms; also configurable). 
When a set of subqueries is generated by a RAG system
and sent to the endpoint individually through asynchronous HTTP requests,
they are usually batched on the server side
to allow them to be processed together by the Engine.
The end user can also simultaneously process multiple top-level queries in a RAG pipeline
and use the batching capability of the retrieval server. 
This exploits the asynchronous nature of HTTP requests. 
With the native support of asynchronous operations in Python,
firing multiple retrieval and LLM requests to an external server
without blocking the program from advancing to other operations until the results are actually needed
is a key ingredient to accelerate the speed of RAG toolkits such as LangGraph,\footnote{https://www.langchain.com/langgraph}
AutoGen,\footnote{https://microsoft.github.io/autogen}
DSPy~\cite{khattab2023dspy}, and
GPT Researcher~\cite{Elovic_gpt-researcher_2023}. 

While batching adds some overhead in gathering queries,
it prevents queries from being processed sequentially.
This results in greater throughput when handling multiple queries. 
This is generally not handled by offline IR toolkits such as
PyTerrier~\cite{PyTerrier} and
Anserini~\cite{anserini},
since online serving is not the primary use case for those tools. 
\routir{} provides the essential wrappers to serve retrieval models,
including those supported by PyTerrier and Anserini,
to efficiently embed them in a RAG system pipeline. 

Furthermore, processors also cache retrieval results
to prevent duplicate requests to the Engine instance. 
\routir{} supports both in-memory and Redis caches,
which provides flexibility to support different cache integrity needs. 

\subsection{On-demand Pipeline Construction}

\begin{figure}[t]
\centering
\begin{json}
{
  "pipeline": "{qwen3-neuclir,plaidx-neuclir}RRF%
  "collection": "neuclir",
  "query": "where is Taiwan"
}
\end{json}
\caption{Example Pipeline Request.
The pipeline issues the query to \texttt{qwen3-neuclir} and \texttt{plaidx-neuclir} engines,
fuses the results with reciprocal rank fusion,
takes the top 50 documents from the fused result,
and finally reranks using Rank1~\cite{weller2025rank1testtimecomputereranking} reranker. }\label{fig:pipeline-request}
\end{figure}

\begin{table}[t]
\setlength\tabcolsep{0.45em}
\caption{Operators for pipeline construction string. Please refer to \url{https://github.com/hltcoe/routir/blob/main/src/routir/pipeline/parser.py\#L7} for the full context-free grammar. }\label{tab:syntax}
    \centering
    \begin{tabular}{l|l|p{7.2cm}}
    \toprule
    Operator            &          Operation  &  Description \\
    \midrule 
    \texttt{e1 >> e2}   &                Pipe &  Pass the retrieval results of \texttt{e1} to a downstream engine \texttt{e2}, such as a reranker. \\
    \texttt{e1\%k}      &               Limit &  Only retain the top $k$ retrieved documents from \texttt{e1} \\
    \texttt{\{e1,
    e2}   &  Parallel Pipelines &  Pass the upstream results or query (if at the beginning of a pipeline) to a list of parallel pipelines (\texttt{e1} and \texttt{e2}). \\
    \texttt{xx\{e1,e2} &    Query Generation &  Generate multiple sub-queries with method \texttt{xx} and issue them to all parallel pipelines. \\
    \texttt{\}xx}       &       Result Fusion &  Fuse retrieval results from parallel pipelines with method \texttt{xx}.  \\
    \bottomrule
    \end{tabular}
    
\end{table}

In addition to serving the query with a single retrieval model,
\routir{} supports on-demand pipeline construction from the end user request.
This is illustrated in Figure~\ref{fig:pipeline-request},
where the pipeline combines the results of two first-stage retrievers
using reciprocal rank fusion and reranks the fused results. 
\routir{} parses the pipeline string provided by the user. 
It understands that the Engines corresponding to the two first-stage retrievers can be run in parallel
and that fusion and reranking are sequential steps with the \texttt{>>} pipe operator. 
In addition, asynchronous requests are issued
to prioritize throughput when producing the final retrieval results.

The pipeline string is defined using a context-free grammar
that supports the construction of linear pipelines.
Table~\ref{tab:syntax} summarizes the operators available for the pipeline construction string. 
Dynamic pipeline construction allows a user or RAG system to control the pipeline as needed on the fly
to accommodate runtime constraints such as latency, coverage, and query difficulty.
The context-free grammar can even be part of the input to the language model,
enabling it to generate the pipeline as part of an agentic workflow.

\section{Serving Models using \routir}

In this section, we describe how \routir{} can be configured and extended to serve different retrieval models. 
This serves as an introduction to all the features provided by \routir{};
readers are encouraged to explore further in the documentation and the source code on GitHub. 

\subsection{Resource and Setup}
Computing resources needed to run the barebone \routir{} are very minimal, for example, a single processor with 200 MB memory can host a \routir{} instance with only Relay Engine. 
However, resources for hosting each retrieval model depend on its own requirements. For example, it is more efficient to host a dense retrieval model with a GPU for encoding the queries; Faiss indexes usually require a larger system memory to hold the in-memory index for better performance. 

\routir{} can be installed through \texttt{pip} or \texttt{uv}. Please refer to the documentation for more details. It can also be hosted with a command as simple as 
\texttt{uvx routir config.json} without explicit package installation. 

\subsection{Server Configuration}\label{sec:usage:config}

\begin{figure}[t]
\centering
\begin{json}
{
  "server_imports": [ "http://localhost:5000" ]
  "file_imports": [ "./examples/rank1_extension.py" ],
  "services": [
    {
      "name": "rank1",
      "engine": "Rank1Engine",
      "config": {}
    }
    {
      "name": "qwen3-neuclir",
      "engine": "Qwen3", 
      "batch_size": 16,
      "config": {
        "index_path": "hfds:routir/neuclir-qwen3-8b-faiss-PQ2048x4fs",
        "embedding_model_name": "Qwen/Qwen3-Embedding-8B",
      }
    }
  ],
  "collections": [
    {
      "name": "neuclir",
      "doc_path": "./neuclir.collection.jsonl"
    }
  ]
}
\end{json}
\vspace{-3mm}
    \caption{Example configuration file.
    Refer to the \routir{} documentation for more fields.
    In the example, we load two engines:
    \texttt{Rank1} reranker from a custom script (see Section~\ref{sec:usage:integration}), and
    \texttt{qwen3-neuclir} using a FAISS index loaded from Huggingface Datasets containing Qwen3 8B embeddings.
    The specified document collection can be used to retrieve document text via an API
    or to pass documents to reranking models.}
    \label{fig:config-example}
\end{figure}

\routir{} uses a JSON file to express the configuration for the services with two primary blocks: 
\texttt{services} and \texttt{collections}. 
The \texttt{services} block is a list of dictionaries
that specifies all the Engines to initialize on the endpoint.
The \texttt{collections} block specifies a list of document collections that the endpoint will serve
based on the document IDs requested. 
The additional top-level \texttt{server\_imports} and \texttt{file\_imports} fields
allow the user to specify external \routir{} endpoints
and custom Python scripts to include during the initialization. 
All available services on each endpoint in \texttt{server\_imports} are automatically relayed. 
This allows offloading computationally expensive models, such as LLM reranking,
to another machine, while still enabling their integration with other services
in the end user's custom retrieval pipelines. 
Figure~\ref{fig:config-example} demonstrates an example configuration JSON object. 

Each dictionary in the \texttt{services} list defines a processor and its underlying Engine. 
The field \texttt{engine} specifies the Engine class to initiate,
which can be any of those included in \routir{} as built-in engines such as \texttt{PLAIDX},
or ones that are implemented by the user in a separate Python script.
This is accomplished via the \texttt{file\_imports} field
that allows users to specify scripts to load on-demand
without modifying the package or implementing another custom entry script.

The \texttt{collections} field lists each collection serviced by an engine
in a dictionary containing the name of the collection and a path to a JSONL file.
Each JSON object is a document containing an ID field with arbitrary fields carrying its content.
\routir{} loads each JSONL collection file and builds a memory offset lookup table to efficiently look up the document when serving the content. 
Such lookup tables allow \routir{} random access the collection file based on document IDs
without reading the file sequentially or loading the entire collection into memory. 

\routir{} provides several built-in Engine types that can be used to serve models with common architectures such as 
dense bi-encoders with FAISS indices,
multi-vector dense retrieval with PLAID-X~\cite{yang2024translate}, and
learned-sparse retrieval models with Anserini~\cite{anserini}. 
These Engines are implemented based on an \texttt{Engine} Python abstract class
that ensures a common interface. 

These Engine types cover most use cases when using neural models. 
However, since there is not yet a common reranker architecture
(besides the general pointwise, pairwise, listwise, and setwise paradigms),
we have selectively implemented a few rerankers as built-in Engines
with an example script for loading a more complex model through \texttt{file\_imports}.\footnote{ Rank1~\cite{weller2025rank1testtimecomputereranking}, a pointwise reasoning reranker, integration script can be found in \url{https://github.com/hltcoe/routir/blob/main/examples/rank1_extension.py}.}
In the next section, we describe how to incorporate an external IR toolkit into \routir. 

\subsection{Integration with Existing IR Toolkits}\label{sec:usage:integration}

\begin{figure}[t]
    \centering
\begin{python}
class PyseriniBM25(Engine):
    def __init__(self, name: str = None, config=None, **kwargs):
        super().__init__(name, config, **kwargs)
        self.searcher = LuceneSearcher(self.index_path)
        self.searcher.set_bm25(0.9, 0.4)

    async def search_batch(self, queries, limit=20):
        return [
            {docobj.docid: docobj.score 
             for docobj in self.searcher.search(query, k=lm)} 
            for query, lm in zip(queries, limit)
        ]
\end{python}
\caption{Example code snippet for integrating Pyserini with \routir.
This example can be extended with more flexible parameter configuration
or even allowing the endpoint users to specify the retrieval model.
Full example can be found at \url{https://github.com/hltcoe/routir/blob/main/examples/pyserini_extension.py}.}
\label{fig:pyserini-example}
\end{figure}

Figure~\ref{fig:pyserini-example} demonstrates a simple example
that integrates Pyserini~\cite{pyserini} into \routir{}. 
For index retrieval, one only needs to implement the initialization of the Engine,
which contains index loading and hyperparameter settings if applicable,
and the \texttt{search\_batch} method,
which takes a batch of queries and returns a list of results in the same order as the input queries. 

Since all search methods are asynchronous,
\routir{} does not wait for a module to finish searching before attending to the next API request.
However, since Python asynchronous functions are still single-threaded on a single processor,
unless the Engine spawns another search process or calls out to another process,
processes can be blocked.
For example, a standalone Lucene instance searching a batch of queries
may block the Python process from accepting an API request.
We provide some implementation guidance on how these concurrency issues can be overcome in the documentation. However, such engineering issues are model and toolkit-dependent.
They can generally be solved by hosting different models in separate \routir{} instances
and combining them via \texttt{server\_imports} to a joint endpoint.

\section{Experiment and Analysis}

To demonstrate the adaptability of \routir{},
we report effectiveness and efficiency using the TREC 2023 NeuCLIR MLIR task~\cite{lawrie2023overview},
which has 76 queries and about 10 million web documents in Chinese, Persian, and Russian
extracted from CommonCrawl News. 
We experiment with the following three multilingual models with distinct architectures and stacks: 
\begin{itemize}
    \item Multi-vector dense using PLAID-X~\cite{yang2024translate}.
    The PLAID-X model was reported as the state of the art in 2023 during TREC.
    The model is based on XLM-RoBERTa-Large~\cite{roberta}
    and is served using an NVIDIA TITAN RTX 24G with the PLAID-X implementation. 

    \item Learned-sparse retrieval using MILCO~\cite{nguyen2025milco} with Anserini~\cite{anserini}.
    The MILCO model is also based on XLM-RoBERTa-Large
    and is served with the same GPU using the Huggingface Transformer.
    The index is served with Anserini via PyJNIus. 

    \item Dense retrieval using Qwen3 Embedding~\cite{zhang2025qwen3} with a FAISS index~\cite{douze2024faiss}
    using vLLM.\footnote{\url{https://github.com/vllm-project/vllm}}
    The Qwen3 Embedding model is served with vLLM with parameters cast to FP16 to fit on the TITAN RTX GPU.
    The document embeddings are indexed with FAISS using product quantization of 2048 dimensions
    and 4-bit fast scan (\texttt{PQ2048x4fs}). 
\end{itemize}

We experimented with two request modes: batched and sequential. 
When queries are batched, we issue all 76 queries with asynchronous HTTP requests to the \routir{} endpoints
and report the throughput, i.e., the number of queries processed per second. 
In the sequential mode, we issued the next query after receiving results from the previous one
 and report the latency, i.e., the number of seconds to process each query. 

\begin{table}[t]
    \setlength\tabcolsep{0.45em}
    \centering
    \caption{Effetiveness and efficiency on the NeuCLIR 2023 MLIR Task. }\label{tab:exp}
    \begin{tabular}{l|c|cc}
    \toprule
    {}                                & Effectiveness  &  Batched Throughput &  Seq. Latency \\
    Model                             &       nDCG@20  &     (query/sec) $\uparrow$  &   (sec/query) $\downarrow$ \\
    \midrule
    Multi-Vector: PLAID-X~\cite{yang2024translate}   &     0.402      &              7.05   &          0.24 \\
    LSR: MILCO~\cite{nguyen2025milco}              &     0.413      &              3.27   &          2.46 \\
    Dense: Qwen3 Embedding~\cite{zhang2025qwen3}  &     0.430      &              9.60   &          1.23 \\
    \bottomrule
    \end{tabular}
    
\end{table}

As shown in Table~\ref{tab:exp}, all three models demonstrate strong throughput,
although LSR with Anserini is the slowest. 
However, it can be greatly accelerated with tools such as Seismic~\cite{bruch2024seismic}
that are tailored for LSR searching. 
All three models are able to take advantage of batched queries to provide faster overall speed. 
Particularly in FAISS, \routir{} uses the batched search capability
to search the index for all queries in the batch at once,
which has the highest throughput despite being five times slower than PLAID-X in sequential latency.

These results demonstrate the adaptivity of \routir{} and robustness to the input types. 
In the next section, we demonstrate how to integrate \routir{} with a RAG system.

\begin{figure}[t]
    \centering
\begin{python}
class LiveRAG_PLAIDX_Search():
    def __init__(self, query: str, **kwargs):
        self.url = os.getenv("RETRIEVER_ENDPOINT")
        self.query = query
        
    def get_content(self, collection, doc_id):
        return requests.post(self.url+"/content", json={
            "collection": collection, "id": doc_id
        }).json()

    def search(self, max_results: int = 5):
        response = requests.post(f"{self.url}/query", json={
            "service": "plaidx-liverag",
            "query": str(self.query), "limit": max_results
        }).json()["result"]

        return [
            {"score": score, "href": str(doc_id),
             **self.get_content("liverag", doc_id) }
            for doc_id, score in results.items():
        ]
\end{python}
\caption{Example search module for GPT Researcher.}
    \label{fig:gptr-example}
\end{figure}

\section{Example of Using \routir{} in a RAG System}

GPT Researcher is an open-source RAG toolkit
that integrates various search engines such as Google and DuckDuckGo into a RAG system
that  supports planning, self-reflections, and multi-agent pipelines. 
Figure~\ref{fig:gptr-example} depicts a retriever in GPT Researcher interfacing with \routir{},
requiring only 21 lines of code. 
Since \routir{} uses the HTTP API as the interface,
it does not require any other dependencies to interface with GPT Researcher. 
Figure~\ref{fig:gptr-example} is an abbreviated version of the implementation
(excluding some sanity checks and try-except blocks) 
used in HLTCOE's system for the LiveRAG challenge at SIGIR 2025~\cite{duh2025hltcoe-liverag}.
While it would also be possible to implement GPT Researcher inside an academic IR platform,
doing so would require substantially more engineering effort.

\section{Summary}

In this work, we introduced \routir,
a simple and robust toolkit for wrapping and serving retrieval models to serve online queries. 
We described the design principles and architecture of \routir{}
and presented several use cases both in configuring the service and integration with RAG pipelines.
This highlighted its ability to increase query throughput,
a particularly desirable feature for RAG systems. 
\routir{} has demonstrated its effectiveness and reliability
as the backbone for the 2025 JHU SCALE Workshop for more than 50 researchers
and also as the search service for 2025 TREC RAGTIME Track. 

\routir{} is still under development with more features in the near term planned,
including integration of LLM rerankers,
Model Context Protocol (MCP) interface,
and better resource management. 
\routir{} is completely open-sourced on GitHub
and welcomes community feedback, feature requests, and pull requests.

\bibliographystyle{splncs04nat}
\bibliography{bibio}

\end{document}